\documentstyle[12pt,aaspp4]{article}
\begin{document}
\title{HST Imaging of ${\bf z > 0.4}$ Quasar Host Galaxies Selected 
by Quasar Radio and Optical Properties\footnotemark[1]}
\footnotetext[1]{
Accepted for publication in the Astrophysical Journal Letters.  
This work is based on observations with the NASA/ESA Hubble Space
Telescope, obtained at the Space Telescope Science Institute, which is
operated by the Association of Universities for Research in Astronomy,
Inc., under NASA contract NAS 5-26555}

\author{Eric J. Hooper\footnotemark[2], Chris D.
Impey\footnotemark[2], and Craig B. Foltz\footnotemark[3] } 
\footnotetext[2]{Steward Observatory, The University of Arizona, Tucson,
AZ 85721 \\ ehooper@as.arizona.edu, cimpey@as.arizona.edu}
\footnotetext[3]{Multiple Mirror Telescope Observatory, The University of
Arizona, Tucson, AZ 85721 \\ cfoltz@as.arizona.edu}

\begin{abstract}
A sample of 16 quasars selected from the Large Bright Quasar Survey in
the redshift range $0.4 < z < 0.5$ has been imaged in the $R$ band
with the Planetary Camera on the WFPC2 instrument of the Hubble Space
Telescope.  The host galaxy magnitudes are mostly similar to or
brighter than $L^*$, and the host luminosity is positively correlated
with the luminosity of the quasar nuclear component.  There is no
distinction in host galaxy magnitude between radio-loud and
radio-quiet quasars, assuming they are all of the same galaxy type.
Many of the host galaxies in the sample have small axial ratios, which
may indicate that they are inclined disk systems.  Alternatively, this
elongated appearance may be due to bars or other distinctive
morphological features which are visible while the bulk of the
underlying lower surface brightness components of the host galaxy are
not.

\end{abstract}


\clearpage

\section{Introduction}
\label{sec:introduction}

The refurbished Hubble Space Telescope (HST) has provided a
substantial boost to the study of quasar host galaxies because of its
ability to resolve the inner structures of the hosts close to the
quasar nucleus.  Most quasars imaged to date with HST lie at redshifts
$z < 0.3$ and have low to moderate nuclear source luminosities
(Bahcall, Kirhakos, \& Schneider 1994\markcite{Bahcall94_ApJ435},
1995a\markcite{Bahcall95_ApJ447}, 1996\markcite{Bahcall96_ApJ457};
Hutchings et al. 1994\markcite{Hutchings94_ApJ429}; Boyce et
al. 1997\markcite{Boyce97}), although Bahcall, Kirhakos,
\& Schneider\markcite{Bahcall95_ApJ450} (1995b) and Disney et 
al.\markcite{Disney95} (1995) include objects with $M_V \approx -27$ in the
cosmology adopted for this paper (H$_0 = 50$ km s$^{-1}$ Mpc$^{-1}$,
q$_0 = 0.5$).  The detected host galaxies are typically at least as
bright as the Schechter function's characteristic luminosity $L^*$,
but several undetected hosts have estimated upper limits of $\leq L^*$
(Bahcall et al.\markcite{Bahcall95_ApJ450} 1995b).  The most recent
determination of $L^*$ in the $R$ passband is $M_R^* = -21.8$ (Lin et
al.\markcite{Lin96} 1996).  While classifying the host galaxies as
either disk or elliptical systems is often difficult because of
contamination by the quasar and the frequent occurrence of disturbed
morphologies, most hosts detected to date in the HST observations
appear to have luminosity distributions consistent with early-type
galaxies (Disney et al.\markcite{Disney95} 1995; Bahcall et
al.\markcite{Bahcall97_ApJ4_97} 1997; and summarized in McLeod \&
Rieke\markcite{McLeod95b} 1995b).  Radio-quiet quasars, which were
once thought to reside in spiral host galaxies (e.g., Smith et
al.\markcite{Smith86} 1986), appear to exist in both elliptical (e.g.,
Hutchings \& Morris\markcite{Hutchings95} 1995) and disk (e.g.,
Hutchings et al.\markcite{Hutchings94} 1994) systems.  Several of the
hosts appear to be involved in interactions or mergers (e.g., Bahcall,
Kirhakos, \& Schneider\markcite{Bahcall95_ApJ447} 1995a), especially
the hosts of infrared luminous quasars (Hutchings \& 
Morris\markcite{Hutchings95} 1995; Boyce et al.\markcite{Boyce97}
1997).

\section{The Sample}
\label{sec:sample}

Candidates for this study were selected from the Large
Bright Quasar Survey (LBQS; Hewett, Foltz, \&
Chaffee\markcite{Hewett95} 1995 and references therein), the largest
homogeneous optical quasar survey to date, containing over 1000
quasars in the redshift range $0.2 < z < 3.4$.  The LBQS was compiled
from digitized UK Schmidt objective prism plates, with candidates
selected based on one or more well-defined selection criteria:
(1) blue color; (2) strong emission or absorption
features; or (3) strong continuum breaks.  Advantages of the survey
include: (1) $> 99$\% success rate in classifying candidates; (2)
sensitivity to almost all known types of quasars, with the exceptions
of the rare classes of objects with red featureless continua or
large-amplitude photometric variations; (3) detection efficiency that
varies smoothly and monotonically with redshift; (4) quantifiable
selection criteria, enabling the calculation of the selection
probability as a function of quasar luminosity, redshift, and spectral
\nopagebreak
energy distribution; and (5) the inclusion of candidates whose images
\nopagebreak
on the discovery plates are resolved due to an underlying host galaxy.
\nopagebreak

This last point is particularly important for studies such as this
one.  Several previous optically selected quasar surveys required that
a candidate appear point-like, a seeing-dependent and often subjective
criterion which could potentially exclude quasars with bright host
galaxies.  Quasars with typical colors and emission line strengths at
the redshifts of the sample selected for WFPC2 imaging satisfy the
LBQS magnitude limit and selection criteria as long as the total
luminosity of the host galaxy is approximately equal to or less than
that of the quasar (Hooper et al.\markcite{radioII} 1995; Hewett et
al.\markcite{Hewett95} 1995).  Since the absolute magnitudes of the
quasars in the HST sample fall in the range $-25 < M_B < -23$, 
very few known galaxies would be excluded {\it a priori}.

The radio properties of the LBQS have been studied using pointed Very
Large Array 8.4 GHz snapshots of 1/3 of the sample (359 targets),
reaching a median detection threshold of about 0.3 mJy (Hooper et
al.\markcite{radioIII} 1996).  The fraction of radio-loud quasars (8.4
GHz luminosity $> 10^{25}$ W Hz$^{-1}$) is constant at $\approx 10$\%
within the errors over almost the full redshift range of the survey
and across nearly the entire absolute magnitude range ($-28 < M_B <
-23$).  The only changes occur at the brightest absolute magnitudes,
$M_B < -28$, where the radio-loud fraction rises to $\approx 30$\%,
and there is also a slight rise around $z \approx 1$.

Selection of targets for WFPC2 imaging was dictated by several
desiderata: (1) a radio luminosity span typical of optically selected
quasars; (2) nearly identical redshifts to avoid problems of
interpretation due to evolution and to provide a uniform rest-frame
spatial resolution scale; and (3) quasar absolute magnitudes bright
enough to exclude Seyfert galaxies, yet not so bright that the nucleus
completely obscures the host galaxy.  LBQS quasars in the redshift
range $0.4 < z < 0.5$ satisfy these criteria.  All six known LBQS
radio-loud quasars within this redshift interval were selected.  These
objects have radio luminosities ranging up to $3 \times 10^{26}$ W
Hz$^{-1}$.  Ten quasars with 8.4 GHz luminosities $\leq 10^{23.5}$ W
Hz$^{-1}$, well below the radio-loud threshold, were selected to
complete the sample.  The radio-quiet quasars were chosen
semi-uniformly from the absolute magnitude range $-25 < M_B < -23$, an
interval containing nearly all of the LBQS quasars with redshifts $0.4
< z < 0.5$.

\section{Data Reduction and Analysis}
\label{sec:data}

The quasars were centered on the Planetary Camera (0.046 arcseconds
per pixel) of WFPC2 and imaged through the F675W filter, which
approximates Johnson $R$.  Total integration times of 20 to 30 minutes
were divided into 3 or 4 separate exposures to avoid saturation of the
quasar core and to facilitate cosmic-ray removal.  The gain was set to
14 electrons per DN in all exposures.  Readnoise, 7 electrons on the
PC, and digitization effects dominated photon noise from the
background in all of the images.  The pipeline data reduction products
were used, which are corrected for bias, dark current, and
flat-fielding.

The reduced data were analyzed by cross-correlating two-dimensional
(2-D) galaxy and point spread function (psf) models with the data to
determine the optimal values of host galaxy flux, exponential
scalelength or effective radius, axial ratio, and position angle, as
well as the quasar centroid and flux.  A description of the general
2-D cross-correlation method was presented by Phillipps \&
Davies\markcite{Phillipps91} (1991); Boyce, Phillipps, \&
Davies\markcite{Boyce93} (1993) discussed its application specifically
to quasar host galaxy analysis.  Advantages of this method include
higher accuracy and sensitivity than one-dimensional profile fitting,
accurate determination of the flux of the quasar nucleus even in low
S/N data where the nature of the host galaxy is rather uncertain, and
relative insensitivity to missing data or the presence of other nearby
galaxies.  Two-dimensional models consisting of a psf, generated with
the ``Tiny Tim'' software (Krist\markcite{Krist96} 1996), and either a
disk or elliptical galaxy component were repeatedly cross-correlated
with the region around the quasar in the WFPC2 images while varying
the psf and model galaxy parameters.  The set of parameters which
maximized the cross-correlation function was adopted.  The 2-D
technique was checked by analyzing one-dimensional (1-D) profiles
derived from elliptical isophote fits to the more prominent host
galaxies, cases in which this older method is still expected to be
fairly robust.  The magnitudes from the 1-D technique were typically
$\approx 0.3$ mag brighter than those from the cross-correlation.

Fluxes were converted to Johnson $R$ using Equation (8) in Holtzman et
al.\markcite{Holtzman95} (1995).  This conversion has a color term of
$< 0.15$ mag for typical disk and elliptical galaxy and quasar colors
at $z \approx 0.5$.  Colors for galaxies and quasars, as well as
$k$-corrections for conversion to rest-frame absolute magnitudes, were
obtained from Coleman, Wu, \& Weedman\markcite{Coleman80} (1980) and
Cristiani \& Vio\markcite{Cristiani90} (1990), respectively.  Minor
($\leq 0.1$ mag) corrections were applied to the derived magnitudes to
account for charge transfer efficiency effects and for differences in
the flux measurement methods between the calibration of the WFPC2
photometric system and the LBQS data, as described in Holtzman et
al.\markcite{Holtzman95} (1995).

\section{Results}
\label{sec:results}

Numerical results of the analysis are presented in Table
\ref{tab:data}, which lists for each target: (1) source name; (2)
redshift from Hewett et al.\markcite{Hewett95} (1995); (3) 8.4 GHz
luminosity (W Hz$^{-1}$) from Hooper et al. (1995, 1996); (4) and (5)
apparent and absolute $R$ magnitudes for the quasar nuclear component,
respectively; apparent (6) and (7)absolute $R$ magnitudes for the host
galaxy using an exponential disk template; (8) and (9) apparent and
absolute $R$ magnitudes for an $r^{1/4}$ model template; and (10) the
axial ratio of the host galaxy.  The quasar magnitudes are relatively
insensitive to the galaxy template used in the analysis, differing by
typically 2\% or less between the two models; the average of the
derived quasar magnitudes is listed in columns (4) and (5).  The
integrated host galaxy fluxes, however, are typically 0.5 to 1
magnitude fainter for the disk model, and both are listed in the
table.  Axial ratios derived from the two models are identical in most
cases, and the average is listed in the few that are not.

Scaled 2-D psfs, normalized to the flux derived for the nuclear point
source, were subtracted from the images in order to better see the
host galaxies.  The region around each quasar is shown in Fig.  1
before (left column) and after (right column) psf subtraction.  The
stretch was adjusted to show the faint outer isophotes in most cases,
although the hard-copy reproduction quality of these faint features is
poor.  Each pixel corresponds to a projected distance of 310 pc at $z
= 0.45$, the midpoint of the redshift range of the sample, $0.4 < z <
0.5$.  The psf subtracted images contain a discontinuity at the center
where the quasar component has been subtracted.  This is an artifact
of the subtraction and is found after a scaled psf is subtracted from
a stellar image.  The discontinuity often appears as a small negative
feature in the images unless the central surface brightness of the
host galaxy is relatively high.  Apparent companion objects lying very
close to the quasar were included in the displayed field.  Many of the
fields contain extended objects on the PC images which are not shown
here.  The two point-like features close to the quasar 0100$+$0205 are
likely to be artifacts associated with the diffraction spikes.

Rigorous analysis of the uncertainties in the derived parameters and
determination of the limiting magnitude of a detectable host galaxy
are challenging problems, which will be addressed in a subsequent
paper.  However, some information on the reliability of the analysis
technique and detection sensitivity was obtained by applying the 2-D
cross-correlation method to stars found in the Planetary Camera quasar
fields.  Comparison with the quasar results is complicated by the fact
that the stars are typically substantially fainter than the quasars.
The brightest star analyzed, in the field of 1218$+$1734, had $R =
20.02$, 0.5 mag fainter than the faintest quasars in the sample
(0020$+$0018 and 1209$+$1259).  The ``host galaxy'' magnitudes
produced for this star were $R = 21.71$ and $R = 22.23$ for the
elliptical and disk templates, respectively, in each case at least 1.4
mag fainter than the results for the quasars.  One quasar,
1243$+$1701, had a very low axial ratio of 0.1 and an unphysically
small effective radius and exponential scalelength of $\leq 300$ pc.
Either there is little or no detectable flux from the host galaxy, or
the host has an unusual morphology and low surface brightness, for
which the standard templates are inadequate.  Some extended residual
emission is seen in the psf subtracted image of this field (Fig. 1).
The small companion very close to the quasar did not affect the
analysis, since similar results were obtained after it had been
removed from the image.  Until additional analysis is performed, it is
not certain whether a host has been detected for this quasar or others
which have faint extended emission only in the near vicinity of the
quasar nucleus.  Consequently, the discussion that follows will
consider both the entire sample and a subset consisting of the most
clearly visible host galaxies.

\section{Discussion}
\label{sec:discussion}

Absolute $R$ magnitude ($M_R$) of the quasar nuclear component is
plotted against host galaxy $M_R$ derived from $r^{1/4}$ and disk
templates in Fig. 2, with radio-loud and radio-quiet quasars denoted
by filled circles and stars respectively.  The detected host galaxies
are luminous; magnitudes derived from $r^{1/4}$ fits and from many of
the disk fits are brighter than $M_R^* = -21.8$ (Lin et
al.\markcite{Lin96} 1996).  Published values of $L^*$ differ by up to
1 mag, but this does not alter the basic result that the elliptical
fits are mostly brighter than the characteristic luminosity, while the
magnitude distribution derived from disk templates straddles $L^*$.
Note in particular that there is not a dominant population of
low-luminosity host galaxies in the LBQS sample.  Host galaxy
luminosity is positively correlated with quasar nuclear luminosity at
the 99\% confidence level for the $r^{1/4}$ and disk template
magnitudes, as indicated by a Kendall's $\tau$ test.  This correlation
is consistent with the existence of a minimum host galaxy luminosity
which increases with quasar luminosity, a trend noted by McLeod \&
Rieke\markcite{McLeod95a}\markcite{McLeod95b} (1995a,b) using $H$-band
host galaxy magnitudes in lower-redshift samples.  The lower bound to
the host galaxy luminosities, converted to $R$-band, is about 0.5
(disk template) or 1.5 (elliptical template) mag fainter than the
lower luminosity envelope of the LBQS host galaxies.  The host galaxy
magnitudes deviate from this trend, becoming roughly constant with
quasar magnitude, for quasars fainter than $M_R \approx -24$.  A
similar break occurs at roughly the same magnitude, after adjusting
for differences in passband and cosmology, in the McLeod \&
Rieke\markcite{McLeod95a}\markcite{McLeod95b} (1995a,b) data.  The
radio-loud quasars all lie in the brighter half of the range of quasar
nuclear $M_R$.  This was not an effect of the selection of the
radio-loud quasars, as the majority were in the fainter half of the
range of $M_B$.  There is no apparent distinction between the host
galaxy magnitudes of radio-loud and radio-quiet quasars with similar
nuclear component luminosities for a single type of galaxy template.
However, if the host galaxy morphologies differ between radio
luminosity classes, such that radio-loud quasars are in ellipticals
and radio-quiet nuclei reside in spirals, then the hosts of luminous
radio sources would be consistently brighter by about one magnitude.

The axial ratios derived for the host galaxies have a median of 0.5,
with only two objects having ratios $> 0.6$.  These values are
significantly lower than is typical for elliptical galaxies, which
generally appear less elongated.  A Kolmogorov-Smirnov (KS) test
indicates that the distribution of axial ratios in a sample of 171
elliptical galaxies from Ryden\markcite{Ryden92} (1992) differs from
the LBQS results at a confidence level $> 99.99\%$.  Fewer than 5\% of
the ellipticals have axial ratios $< 0.6$, where the bulk of the LBQS
objects lie.  This result may indicate that the LBQS sample consists
primarily of inclined disk systems, which differs from the conclusion
of several other HST imaging studies that early type host galaxies are
prevalent, even among radio-quiet objects.  McLeod
\& Rieke\markcite{McLeod94a}\markcite{McLeod94b} (1994a,b) did not
find any host galaxies with measurable axial ratios $< 0.5$ in a
sample drawn from the PG quasar survey, even among galaxies which
appeared to be spirals.  They attributed the lack of inclined disk
systems to the additional reddening of the active nucleus caused by
this viewing geometry, which would cause the quasar to fail the
UV-excess selection criterion of the PG.  Increased extinction would
also remove potential candidates by making them fainter than the PG
magnitude limit.  However, very few quasars which are intrinsically
too bright to be included in the PG would cross the magnitude
selection boundary of the survey due to extinction from an inclined
host, because the bright flux limit of the PG exceeds the apparent
fluxes of virtually all known quasars.  The LBQS selection is based on
spectral features in addition to color, and its bright magnitude limit
is roughly coincident with the faint limit of the PG, so it is
reasonable to expect that spiral host galaxies with higher
inclinations could be included in the LBQS.  Another possibility is
that the detected portions of some of the host galaxies are bars,
particularly in the cases of very low axial ratios.  These systems
could have nearly face-on disks which are too faint to detect, leaving
only the bar components visible.

Half of the sample, the 8 targets with the most obvious host galaxies
in Fig. 1, was examined separately to ensure that the major results
are not skewed by the lower surface brightness host galaxies, for
which only marginal emission is seen and which may not have been
detected in some cases.  This subsample contains 3 radio-loud quasars
(1222$+$1235, 1230$-$0015, and 2348$+$0210) and 5 radio-quiets
(0020$+$0018, 0021$-$0301, 1149$+$0043, 1222$+$1010, 2214$-$1903).
The absolute magnitudes of the quasar nuclear components in these
objects span the range of the full sample.  A correlation with host
galaxy luminosity is still suggested visually, particularly among the
more luminous quasars, but the trend is too week given the reduced
sample size to produce a significant result from the Kendall's $\tau$
test (86\% confidence).  Radio-loud and radio-quiet quasars with
similar central component magnitudes in the subsample have similar
host galaxy luminosities when derived from the same class of galaxy
template.  Only one host galaxy in the restricted sample has an axial
ratio $> 0.6$, and the flattened morphologies of the remaining objects
can be easily seen in most cases in Fig.  1.  The KS test using
Ryden's (1992) sample of elliptical galaxies was repeated, with the
conservative assumption that the host galaxies not included in the
subsample had relatively round morphologies closely matching the axial
ratio distribution of the ellipticals.  The two distributions were
different at $> 99\%$ confidence level, indicating that there is a
significant component of clearly flattened systems in the LBQS sample.

\acknowledgments

The authors thank Alice Quillen, Gary Schmidt, and Joe Pesce for
helpful discussions and the referee, Donald Schneider, for
constructive input.  This work was supported by NASA through GO
program grant number 5450 from the Space Telescope Science Institute,
which is operated by the Association of Universities for Research in
Astronomy, Inc., under NASA contract NAS 5-26555.  Additional support
was provided by a NASA Graduate Student Researchers Program Fellowship
(EJH), grant number NGT-51152.  This research has made use of the NASA
Astrophysics Data System (ADS).

\clearpage

\section{Figure Captions}
\label{sec:captions}

\figcaption{Each pair of linearly stretched grayscale
images across the page shows the field around one of the quasars in
the sample.  The quasar and its host galaxy are shown in the left
panel, and the same field after the nuclear component has been
subtracted is displayed in the right panel.  A direction indicator
(the arrow points north and the other line east) and size scale are
printed between each pair of panels.  The LBQS designation of
the quasar is listed above each image pair.  These are only the
centers of the WFPC2 PC frames, sized to include all detectable
emission from the host galaxies and any particularly close apparent
companions.}

\figcaption{Host galaxy absolute magnitude derived
from (a) elliptical and (b) disk templates plotted against absolute
magnitude of the quasar nuclear component.  Filled circles are
radio-loud quasars, and asterisks represent radio-quiet sources. }

\clearpage

\begin{deluxetable}{lccccccccc}
\scriptsize
\tablenum{1}
\tablewidth{0pt}
  
\tablecaption{\label{tab:data} Results of WFPC2 Analysis}

\tablehead{
\colhead{\parbox{1cm}{\centering   Name \phn \\ (1)  }} &
\colhead{\parbox{1cm}{\centering   $z$ \\ \vspace{5mm} (2)  }} &
\colhead{\parbox{2cm}{\centering   $\log L_{8.4}$ \\ (W Hz$^{-1}$) \\ (3) }}  &
\colhead{\parbox{1.5cm}{\centering  $R$ \\ (quasar) \\ (4) }}  &
\colhead{\parbox{1.5cm}{\centering   $M_{R}$ \\ (quasar)  \\ (5) }}  &
\colhead{\parbox{1.0cm}{\centering   $R$ \\ ($r^{1/4}$)  \\ (6) }}  &
\colhead{\parbox{1.0cm}{\centering   $M_{R}$ \\ ($r^{1/4}$)  \\ (7) }} & 
\colhead{\parbox{1.0cm}{\centering   $R$ \\ (disk)  \\ (8) }}  &
\colhead{\parbox{1.0cm}{\centering   $M_{R}$ \\ (disk)  \\ (9) }} &
\colhead{\parbox{1.0cm}{\centering   axial \\ ratio \\ (10) }} 
}
\startdata

0020$+$0018 & 0.423 & $<$ 23.50 & 19.52 & $-$22.39 & 19.33 & $-$23.42 & 20.17 &
$-$22.19 & 0.7 \\
0021$-$0301 & 0.422 & $<$ 23.50 & 19.15 & $-$22.75 & 19.87 & $-$22.87 & 20.67 &
$-$21.68 & 0.6 \\
0100$+$0205 & 0.393 & $<$ 23.45 & 17.81 & $-$23.93 & 19.97 & $-$22.54 & 20.70 &
$-$21.47 & 0.3 \\
1138$+$0003 & 0.500 & $\phm{<}$ 25.60  & 17.94 & $-$24.35 & 19.33 & $-$23.97 & 20.26 &
$-$22.54 & 0.9 \\
1149$+$0043 & 0.466 & $<$ 23.66  & 17.37 & $-$24.76 & 19.17 & $-$23.91 & 19.77 &
$-$22.84 & 0.6 \\
1209$+$1259 & 0.418 & $<$ 23.52  & 19.56 & $-$22.32 & 20.20 & $-$22.51 & 20.85 &
$-$21.48 & 0.6 \\
1218$+$1734 & 0.445 & $\phm{<}$ 25.33  & 18.36 & $-$23.66 & 19.63 & $-$23.28 & 20.33
& $-$22.15  & 0.6 \\
1222$+$1010 & 0.398 & $<$ 23.46  & 18.46 & $-$23.30 & 19.36 & $-$23.18 & 20.59 &
$-$21.60 & 0.3 \\
1222$+$1235 & 0.412 & $\phm{<}$ 25.19  & 17.86 & $-$23.98 & 18.85 & $-$23.81 & 19.55 &
$-$22.74 & 0.6 \\
1230$-$0015 & 0.470 & $\phm{<}$ 25.77  & 17.51 & $-$24.64 & 18.96 & $-$24.14 & 19.65 &
$-$22.98 & 0.4 \\
1240$+$1754 & 0.459 & $<$ 23.47  & 18.13 & $-$23.96 & 19.50 & $-$23.50 & 20.40 &
$-$22.16 & 0.3 \\
1242$-$0123 & 0.491 & $<$ 23.47  & 18.26 & $-$23.99 & 19.51 & $-$23.73 & 19.91 &
$-$22.84 & 0.4 \\
1243$+$1701 & 0.459 & $<$ 23.41  & 18.71 & $-$23.38 & 20.84 & $-$22.24 & 20.92 &
$-$21.64 & 0.1 \\
2214$-$1903 & 0.397 & $<$ 23.48  & 18.98 & $-$22.77 & 19.18 & $-$23.35 & 19.75 &
$-$22.43  & 0.5 \\
2348$+$0210 & 0.504 & $\phm{<}$ 25.56  & 17.18 & $-$25.13 & 19.57 & $-$23.75 & 20.25 &
$-$22.57 & 0.4 \\
2351$-$0036 & 0.460 & $\phm{<}$ 26.40  & 17.92 & $-$24.18 & 20.07 & $-$23.02 & 20.87 &
$-$21.70 & 0.4 \\

\enddata
\end{deluxetable}

\clearpage

\end{document}